\documentclass{article}

\usepackage{arxiv}

\usepackage[utf8]{inputenc} 
\usepackage[T1]{fontenc}    
\usepackage{hyperref}       
\usepackage{url}            
\usepackage{booktabs}       
\usepackage{amsfonts}       
\usepackage{nicefrac}       
\usepackage{microtype}      
\usepackage{lipsum}
\usepackage{float} 
\usepackage{graphicx}
\graphicspath{ {./images/} }

\title{Multi‑Layer Protection Against Low‑Rate DDoS Attacks in Containerized Systems}

\author{
 Ahmad Fareed \\
  EPITA \\
  School of Engineering and Computer Science\\
  Le Kremlin-Bicêtre, 94270, France \\
   \And
 Bilal Al Habib \\
  EPITA \\
  School of Engineering and Computer Science\\
  Le Kremlin-Bicêtre, 94270, France \\
  \And
 Anne Pepita Francis \\
  EPITA \\
  School of Engineering and Computer Science\\
  Le Kremlin-Bicêtre, 94270, France \\
}

\begin{document}
\maketitle
\begin{abstract}
Low-rate Distributed Denial-of-Service (DDoS) attacks have emerged as a major threat to containerized cloud infrastructures. Due to their low traffic volumes, these attacks can be difficult to detect and mitigate, potentially causing serious harm to internet applications. This work proposes a DDoS mitigation system that effectively defends against low-rate DDoS attacks in containerized environments using a multi-layered defense strategy. The solution integrates a Web Application Firewall (WAF), rate limiting, dynamic blacklisting, TCP/UDP header analysis, and zero-trust principles to detect and block malicious traffic at different stages of the attack life-cycle. By applying zero-trust principles, the system ensures that each data packet is carefully inspected before granting access, improving overall security and resilience. Additionally, the system’s integration with Docker orchestration facilitates deployment and management in containerized settings. 
\end{abstract}

\section{Introduction}
In today's fast-changing digital landscape, many organizations are adopting container-based cloud setups to improve how they manage applications, making operations more flexible, scalable, and efficient. However, this shift introduces new challenges, especially in maintaining security. One major challenge is the rise of low-rate Distributed Denial of Service (DDoS) attacks, which can disrupt the performance and safety of containerized systems. Unlike high-volume flooding attacks, low-rate DDoS attacks operate at traffic levels close to legitimate usage, by exploiting protocol behavior and application-layer characteristics, making them particularly difficult to detect using traditional volume-based defense mechanisms. \cite{low-rate_DDOS}

This research focuses on finding practical solutions to tackle these low-rate DDoS attacks in container-based cloud environments. The main goal is to help organizations strengthen their defenses against evolving cyber threats while ensuring the integrity of their web applications. Our aim is to develop a robust security strategy that protects organizations from persistent cyber-attacks and maintains the uninterrupted operation of their applications by detecting DDoS activity in containerized systems.

\section{Motivations}
The zero-trust model embraced in this project is motivated by a paradigm shift in cybersecurity, challenging traditional assumptions about trust within network architectures. Recognizing the escalating threat landscape, particularly the pervasive nature of Distributed Denial of Service (DDoS) attacks, the zero-trust model becomes a pivotal approach for fortifying web applications. Its primary aim is to eliminate any default trust in the network, prompting rigorous scrutiny of all traffic regardless of origin.

Each stage of the defense mechanisms in this work reflects this approach, with rate limiting, dynamic blacklisting, TCP and UDP header analysis, and Web Application Firewall (WAF) deployment contributing to a comprehensive zero-trust architecture. By scrutinizing traffic at every entry point, the system operates under the principle of "never trust, always verify," fortifying web applications against DDoS attacks and other potential threats. \cite{0trust}

This research further emphasizes the significance of micro-segmentation in enhancing security. Micro-segmentation partitions the network into granular segments with unique security parameters, restricting lateral movement and mitigating the potential impact of a breach, aligning seamlessly with the zero-trust philosophy. 

\section{Objective}
The main objective of this work is to develop and implement an innovative real-time function within container-based cloud environments. The choice of a container-based environment is motivated by previous studies that demonstrate how container and microservice architectures offer unique opportunities and constraints for mitigating low-rate DDoS attacks. \cite{Defeat_DDOS_on_Containers} This function will effectively distinguish between normal and malicious traffic, thereby establishing a robust defense mechanism against low-rate Distributed Denial of Service (DDoS) attacks. In achieving this objective, the project will integrate the principles of the Zero Trust Model to enhance the overall security posture of containerized systems, ensuring continuous verification and minimizing the attack surface for increased resilience against cyber threats.
By integrating the Zero Trust Model with the proposed real-time function, the project aims to create a comprehensive and adaptive defense strategy for containerized systems, addressing the unique challenges posed by low-rate DDoS attacks in the cloud environment.

\section{Related Work}
Low-rate DDoS attacks have been studied extensively in both cloud and containerized environments. Previous research includes surveys summarizing detection and defense mechanisms~\cite{low-rate_DDOS}, approaches for container-based mitigation~\cite{Defeat_DDOS_on_Containers}, TCP- and UDP-based attack detection~\cite{tcp_low_rate, udp_ddos_flood}, and zero-trust container architectures~\cite{0trust}. Our work presents a practical, multi-layered defense system that integrates rate limiting, dynamic blacklisting, TCP/UDP header analysis, and WAF deployment within a containerized environment.

\section{Proposed Defense Strategy}
This work presents a multi-layered security strategy for mitigating low-rate DDoS attacks which includes rate limiting, blacklisting, header analysis, and web application firewall (WAF). The subsequent chapters will build upon these insights to devise a robust defense mechanism tailored to the unique challenges posed by low-rate DDoS attacks in the cloud environment.

Rather than relying solely on traffic volume thresholds, the proposed strategy focuses on behavioral analysis, protocol-level inspection, and continuous verification to counter stealthy low-rate attack patterns. 

\subsection{First Line of Defense: Rate Limiting}
The system architecture is composed of two containers using Docker, the initial line of defense for detecting and managing incoming traffic is established through rate limiting. This mechanism is configured in the first container, which handles the ingress traffic.

Rate limiting involves setting predefined thresholds on the number of requests or data packets allowed within a specified time frame. By implementing rate limiting at the entry point of the system, we control the flow of incoming traffic, preventing issues such as traffic spikes, and resource exhaustion. This approach adds a layer of security, enhancing the overall stability, reliability, and performance of the system. The rate-limiting mechanism serves as a proactive measure, ensuring the system operates within specified capacity limits and remains resilient in the face of potential abuse or unexpected surges in traffic. This strategic use of rate limiting contributes to the effective management and protection of the system, creating a controlled and secure environment for processing incoming requests.

Rate limiting is employed as a form of traffic shaping within the first container, specifically applied to manage the ingress traffic flow. This technique allows us to control the volume and distribution of incoming traffic from the internet, preventing potential overload scenarios that could lead to system failure. Without rate limiting, incoming traffic could accumulate in a queue, causing congestion and adversely affecting the system's performance.

The rate limiting middleware, integrated into the first container handling ingress traffic, ensures a balanced and fair distribution of requests from different sources on the internet. Each source is assigned its own rate limiter within the middleware, and the configuration parameters, such as burst and requests per second, are customizable. This customization lets us define specific limits for each source, preventing it from monopolizing the system's resources.

For instance, we can configure the rate limiting middleware to permit each source a maximum of five requests per second with a burst allowance of ten requests. If a source reaches its capacity, it is temporarily restricted until the rate limiter's "bucket" is less full, allowing for a more equitable distribution of requests.

\subsection{Second Line of Defense: Dynamic Blacklisting}
The Second Line of Defense for the Ingress Traffic Flow is Dynamic Blacklisting. Dynamic blacklisting is a technique used to block malicious endpoints from attacking the network. It is used to prevent Denial of Service (DoS) attacks by monitoring signaling traffic and dynamically detecting potential attacks without disrupting the rest of the services that it provides. The attacks can then be blocked internally or externally.Dynamic blacklists are put in place automatically by the system when it detects an attempt to disrupt traffic flowing through it. Implementing dynamic blacklisting involves actively monitoring incoming traffic for malicious behavior and dynamically updating firewall rules to block or drop traffic from known abusers.

In this system, the suspicious IPs ranges are continuously fetched from the internet and stored in a file. When a new request arrives, the system first checks the IPs file. If the IP address is not on the list, the traffic is allowed. If it is on the list, the traffic is rejected.

\begin{figure}[H]
\centering
\includegraphics[width=0.8\linewidth]{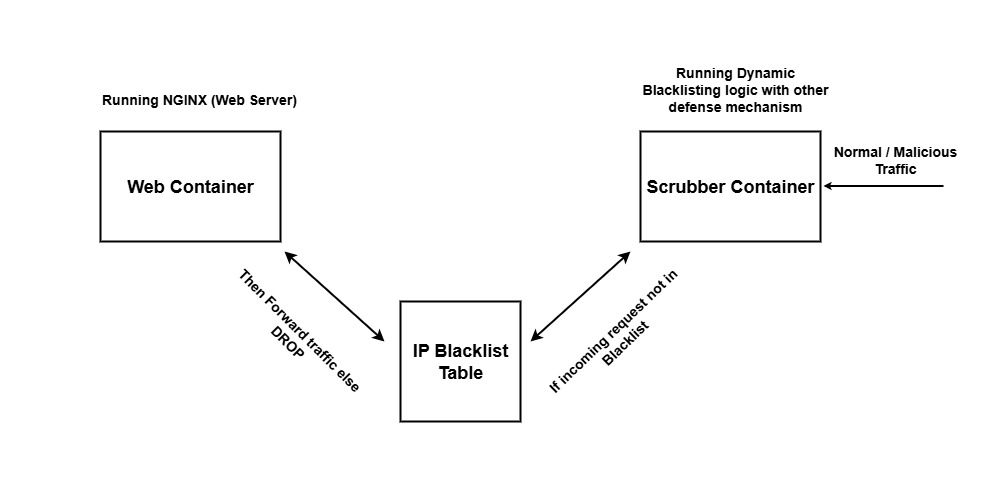}
\caption{Dynamic Blacklisting}
\label{fig:dynamic_blacklisting}
\end{figure}

\subsection{Third Line of Defense: Analyzing the TCP/UDP Packets}
In this project, a foundational approach to thwarting Distributed Denial of Service (DDoS) attacks involves a meticulous examination of TCP and UDP headers, acting as a critical strategy to pinpoint irregularities in network traffic. The continuous monitoring of these headers plays a pivotal role in detecting potential threats, such as SYN or ACK floods, by establishing thresholds for normal traffic patterns and triggering alerts when deviations occur. Specifically, the analysis of TCP headers is instrumental in promptly identifying and implementing countermeasures against various types of attacks. For instance, anomalies like SYN requests without completing the three-way handshake can signify a potential SYN flood attack. To address this, countermeasures such as SYN cookies, rate limiting, and port-based filtering are deployed, enhancing the system's resilience.

Furthermore, the scrutiny of payload content is integral to identifying potentially malicious data within the network traffic. The incorporation of automated responses enhances the system's real-time responsiveness, ensuring effective mitigation of DDoS attacks and fortifying the project's network infrastructure against disruptions.

The third line of defense in our DDoS filtering strategy involves a comprehensive analysis of TCP and UDP flags to identify and mitigate potential threats. Within the TCP protocol, a focused examination of specific flags—such as SYN, ACK, RST, PSH, and URG—provides valuable insights into network traffic patterns. For instance, a high rate of SYN requests without completing the three-way handshake could indicate a SYN flood attack, prompting the implementation of SYN cookies or rate-limiting measures. Similarly, an unusual surge in ACK packets without corresponding data transmission might indicate an ACK flood attack, necessitating the deployment of rate-limiting for ACK packets. Additionally, the analysis considers the frequency of RST packets, especially when accompanied by other irregular activities, as they may suggest attempts to disrupt communication. Proper RST packet handling and rate limiting become crucial in addressing such scenarios. Furthermore, the scrutiny of the uncommon use of PSH flags, potentially indicating attempts to push malicious payloads, is addressed by regularly inspecting and sanitizing incoming data. \cite{tcp_low_rate}

In the UDP protocol, attention is directed toward critical parameters such as packet length, checksums, and source/destination ports. Measures like size-based filtering, checksum validation, and port-based filtering are implemented to prevent and mitigate malicious activities effectively. This meticulous examination of flags and parameters in the third line of defense aims to detect and neutralize anomalous network behavior associated with potential DDoS attacks, thereby significantly enhancing the overall security posture of the system. \cite{udp_ddos_flood}

\subsection{Fourth Line of Defense: WAF – Mod-Security}
The fourth line of defense in this DDoS detection project is the Web Application Firewall (WAF), specifically implemented through Mod-Security. This acts as a critical shield against potential threats targeted at the application layer. This serves as a robust defense mechanism, safeguarding the web application from a spectrum of cyber threats. WAFs, in general, operate by inspecting and filtering HTTP traffic, making them adept at identifying and thwarting malicious activities, including those associated with DDoS attacks, at the application layer. Mod-Security's strength lies in its rule-based engine, enabling the creation and customization of rules tailored to the project's unique requirements. This adaptability proves invaluable in mitigating evolving security challenges. By conducting real-time analysis of HTTP requests and responses, Mod-Security can promptly detect and respond to suspicious patterns, malicious payloads, and various application layer attacks. Its capability to block or log potentially harmful traffic enhances the project's overall resilience, making Mod-Security an indispensable component in fortifying the web application against cyber threats.

\subsection{DDoS Sandbox}
The above-mentioned components identify the DDoS malicious traffic and forward it toward the sandbox. The DDoS Sandbox plays a vital part in the defense architecture. Using network micro-segmentation and the zero-trust paradigm, all the malicious traffic is forwarded to the DDoS sandbox to reduce the risk to the main infrastructure. It protects primary containers against disruption and effectively mitigates emerging threats by isolating and analyzing harmful patterns. To put it simply, the DDoS Sandbox is a proactive step in our defense-in-depth plan that protects our network infrastructure's integrity and resilience from malicious activity.

\subsection{Tools Used}
The following tools were used to implement and test the DDoS mitigation system:

\begin{description}
  \item[Apache2:] Apache2 is a well-known open-source web server software that is both flexible and efficient. It serves both static and dynamic content, making it a key component of modern web infrastructure.

  \item[Docker:] A containerization platform that simplifies application deployment by encapsulating software and its dependencies in lightweight, isolated containers.

  \item[Mod-Security:] An open-source Apache2 web application firewall (WAF) that provides real-time monitoring, logging, and protection against web-based threats using the OWASP Core Rule Set.

  \item[hping3:] A network testing tool used to simulate TCP and UDP SYN-based attack traffic for evaluating system resilience.

  \item[Mausezahn:] A traffic generation tool used to emulate high-volume attack scenarios and assess system behavior under stress.

  \item[Iperf3:] A network performance measurement tool used to generate normal traffic and establish a baseline for comparison.
\end{description}

\section{System Architecture}
The system architecture includes two containers, each of which plays an important role in improving the security and efficiency of the system's operations. The primary container is the core hub for integrating and deploying all DDoS detection technologies. In the meantime, the secondary container isolates and separates detected harmful traffic from the primary web container, acting as a dedicated DDoS sandbox.

The web container, which runs on the Ubuntu platform, houses the Apache2 web server, which is the foundation of the system's web hosting capabilities. To protect the integrity and security of the system, all incoming traffic is carefully screened before entering the system. First, rate-limiting methods are applied to the traffic to control and lessen sudden spikes in traffic that could overload the system. When traffic reaches predetermined levels, it is immediately dropped to avoid congestion and maintain system efficiency.

The traffic is then subjected to dynamic IP blacklisting checks, which compare incoming requests to a constantly updated blacklist of known malicious IP addresses. Requests coming from IP addresses that have been blacklisted are immediately turned down, reducing the possibility of threats and illegal access attempts. On the other hand, legitimate communication is allowed to pass through for additional examination, demonstrating the system's strong defenses against outside threats.

Following the initial screening processes, the traffic is meticulously analyzed to discover any unusual patterns or abnormalities that could suggest hostile intent. Using the capabilities of the robust web application firewall (WAF) Mod-Security, the traffic is carefully examined against the extensive core rule set that OWASP provides. By using a proactive approach to security, a wide range of web-based assaults can be detected and mitigated in real time, strengthening the system's resistance to new threats and weaknesses.

Malicious traffic detected is seamlessly transferred to a specialized sandbox container for additional inspection and isolation, keeping the primary web server from being compromised and assuring uninterrupted service delivery to legitimate customers. On the other hand, harmless traffic is gently directed towards the Apache2 web server, allowing users to access web content and services without difficulty while maintaining strict security guidelines. By means of the careful coordination of containerized elements and strong security measures, the revised system architecture is an example of a proactive and robust strategy for protecting against dynamic cyberattacks and guaranteeing the uninterrupted functioning of web services.

\begin{figure}[H]
\centering
\includegraphics[width=0.9\linewidth]{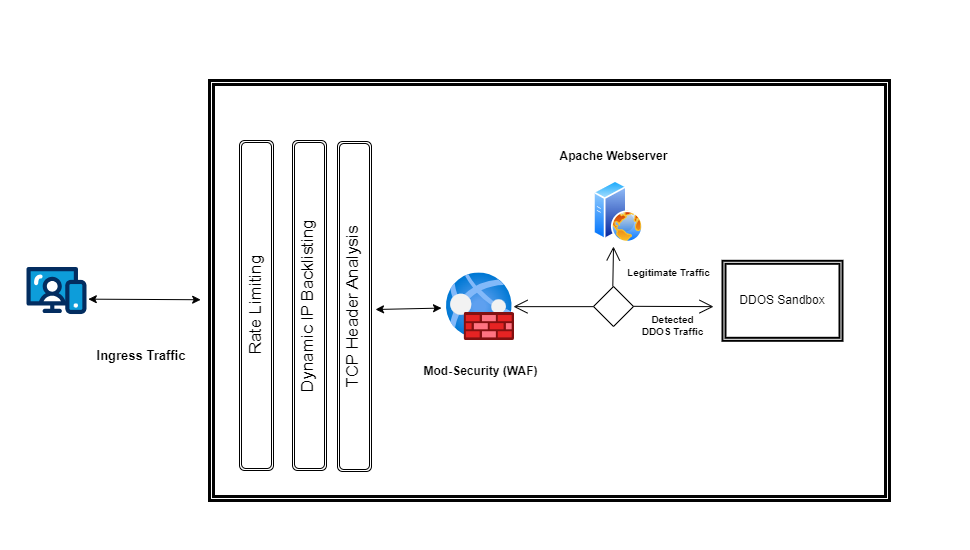}
\caption{Proposed system architecture diagram}
\label{fig:system_architecture}
\end{figure}

\section{Methodology}
The proposed DDoS mitigation system operates through a layered workflow to ensure resilient defense in containerized environments. Initially, incoming traffic is monitored and regulated using rate-limiting to prevent sudden surges from overwhelming the system. Subsequently, requests are cross-checked against dynamic blacklists to block access from known malicious sources.

Traffic is then scrutinized at the network level, analyzing TCP and UDP headers to detect anomalies such as SYN or ACK floods. At the application layer, Mod-Security performs a detailed inspection to identify and block malicious payloads. Traffic deemed harmful is redirected to a dedicated sandbox container, isolating threats from the primary web server.

This stepwise approach enforces continuous verification for all data packets, following zero-trust principles, while allowing legitimate traffic to reach the web application without disruption.

\section{Results}
The proposed DDoS mitigation system efficiently defends web applications against low-rate DDoS attacks in containerized settings. A multi-layered protection mechanism is implemented to enhance the security of the container-based cloud environment. The integration of Mod-Security and the zero-trust model allows the system to manage large volumes of malicious traffic while maintaining the availability of web applications. Additionally, the use of Docker orchestration simplifies deployment and management within containerized systems.

The results demonstrate that the implemented zero-trust model provides a resilient and effective defense against distributed denial-of-service (DDoS) attacks. Traffic passing through the system is subjected to multiple verification stages, ensuring that only legitimate traffic is allowed to reach the Apache2 web server hosted within the web container. This layered validation confirms the effectiveness of the zero-trust principle, where no traffic is implicitly trusted.

All incoming traffic undergoes rigorous inspection as it traverses the system’s defense layers. The first observed result is the effectiveness of rate-limiting mechanisms in regulating traffic flow and reducing the likelihood of congestion caused by abnormal request rates. The second layer, dynamic IP blacklisting, successfully identifies and blocks malicious endpoints, strengthening the system’s resistance to Denial of Service (DoS) attempts.

Further analysis shows that inspecting TCP and UDP header fields enables timely detection of anomalies such as SYN and ACK floods. This capability allows the system to rapidly identify and mitigate sophisticated low-rate DDoS attack patterns, contributing to improved overall network stability.

At the application layer, the deployment of the Mod-Security Web Application Firewall (WAF) proves effective in analyzing HTTP traffic against the OWASP Core Rule Set. This inspection mechanism successfully detects and mitigates a wide range of web-based threats, including application-layer denial-of-service attacks.

Finally, traffic identified as malicious is consistently redirected to the sandbox container, preventing any impact on the primary web server. Legitimate traffic, once validated, is forwarded to the Apache2 service without interruption. These results confirm that the coordinated operation of multiple security mechanisms establishes a strong zero-trust environment, ensuring service availability while effectively mitigating low-rate DDoS attacks. It should be noted that the evaluation presented here is qualitative, based on observed system behavior under low-rate DDoS attack scenarios, rather than quantitative performance metrics.

\section{Conclusion}
The proposed DDoS mitigation system efficiently defends web applications against low-rate DDoS attacks by adopting a multi-layered protection strategy. Rate limiting, dynamic blacklisting, TCP/UDP header examination, a Web Application Firewall (WAF), and zero-trust principles are all used in this technique. Together, these technologies recognize and stop malicious traffic at different phases of the attack lifecycle. Furthermore, the system conforms to zero-trust principles, guaranteeing that each data packet is examined thoroughly prior to authorization. By removing the presumption of trust and demanding ongoing verification, this method improves security. Lastly, deployment and administration in containerized settings are made simpler by the system's connection with Docker orchestration.
The system's efficacy was shown in a containerized environment. A variety of web-based applications were effectively protected against low-rate DDoS attacks using the proposed system. The system was also simple to deploy and maintain, and it was scalable.
While the suggested DDoS mitigation system is effective against low-rate attacks, there remains room for improvement. To cover a wider range of attack methods, the system can be expanded to include honeypots, which attract malicious actors and disclose information about their attack strategies. Furthermore, by automating the identification and classification of malicious traffic, machine learning (ML) techniques can be incorporated into the system, improving its efficiency and efficacy. By making these improvements, the suggested system can attain its full potential and become an even more effective defender against cyber-attacks in the ever-changing digital ecosystem.

\bibliographystyle{unsrt}  
\bibliography{references}

\end{document}